\documentclass[usenatbib,usegraphicx]{mn2e}

\def\nd{--~~~~~~}
\def\md{--~~\,}
\def\sp{~~~~~~~}

\title[Radio Modulations in SN 2001ig]{Modulations in the radio light
curve of the Type~IIb Supernova 2001ig: Evidence for a Wolf-Rayet
binary progenitor?}
\author[S. D. Ryder et al.]{Stuart D. Ryder$^{1}$\thanks{E-mail:
sdr@aaoepp.aao.gov.au}, Elaine M. Sadler$^{2}$, Ravi Subrahmanyan$^{3}$,
Kurt W. Weiler$^{4}$,
\newauthor Nino Panagia$^{5}$ and Christopher Stockdale$^{6,4}$\\
$^{1}$Anglo-Australian Observatory, P.O. Box 296, Epping, NSW 1710, Australia\\
$^{2}$School of Physics, University of Sydney, NSW 2006, Australia\\
$^{3}$Australia Telescope National Facility, CSIRO, Locked Bag 194, Narrabri,
NSW 2390, Australia\\
$^{4}$Naval Research Laboratory, Code 7213, Washington, DC 20375-5320,
U.S.A.\\
$^{5}$ESA/Space Telescope Science Institute, 3700 San Martin Drive, Baltimore,
MD~21218, U.S.A.\\
$^{6}$Physics Dept., Marquette University, P.O. Box 1881, Milwaukee, WI~53201,
U.S.A.}
\begin{document}

\date{Accepted 2004 January 8; in original form 2003 October 14}

\pagerange{\pageref{firstpage}--\pageref{lastpage}} \pubyear{2004}

\maketitle

\label{firstpage}

\begin{abstract}

We describe the radio evolution of SN~2001ig in NGC 7424, from
700~days of multi-frequency monitoring with the Australia Telescope
Compact Array (ATCA) and the Very Large Array (VLA). We find that
deviations of the radio light curves at each frequency from the
standard ``minishell'' model are consistent with density modulations
in the circumstellar medium (CSM), which seem to recur with a period
near 150~days. One possibility is that these are due to enhanced
mass-loss from thermal pulses in an AGB star progenitor. A more likely
scenario however is that the progenitor was a Wolf-Rayet star, whose
stellar wind collided with that from a massive hot companion on an
eccentric 100~day orbit, leading to a regular build-up of CSM material
on the required time and spatial scales. Recent observations of ``dusty
pinwheels'' in Wolf-Rayet binary systems lend credibility to this model.
Since such binary systems are also thought to provide the necessary
conditions for envelope-stripping which would cause the Wolf-Rayet
star to appear as a Type~Ib/c supernova event rather than a Type~II,
these radio observations of SN~2001ig may provide the key to linking
Type~Ib/c SNe to Type~IIb events, and even to some types of Gamma-Ray
Bursts.

\end{abstract}


\section{Introduction}
\label{s:intro}

Radio studies of supernovae (SNe) can provide valuable information
about the density structure of the circumstellar medium, the late
stages of stellar mass-loss, and independent distance estimates for
the host galaxies \citep{kw02}. Furthermore, with the growing
realisation that some Gamma-Ray Bursts may be intimately linked with
SNe [e.g., GRB980425 and SN~1998bw \citep{kul98}; GRB011121 and
SN~2001ke \citep{gar03}; GRB030329 and SN~2003dh \citep{sta03}], such
studies are crucial to understanding GRB environments. To date, radio
emission has only ever been detected from core-collapse SNe of Type II
and Type Ib/c [for a recent review of supernova taxonomy, see
\citet{tur03}], and not at all from thermonuclear Type~Ia SNe.

Supernova~2001ig was discovered visually by \citet{evans01} on 2001
Dec 10.43 UT in the outskirts of the nearby late-type spiral galaxy
NGC~7424 ($D=11.5$~Mpc; \citet{tu88}). No supernovae have been recorded
previously in this galaxy. Initial optical spectroscopy of SN~2001ig
from Las Campanas Observatory \citep{mj01} revealed similarities to
the Type~IIb SN~1987K \citep{fil88}, while spectra from the European
Southern Observatory over the following month \citep{cp01,clo02}
showed a similar behaviour to that of the Type~IIb SN~1993J, by
transitioning from Type~II to Type Ib/c as the H recombination lines
weakened. By October 2002, the transition to a Type Ib/c in the
nebular phase was well and truly complete \citep{fc02}. SN~2001ig was
also detected by the ACIS-S instrument on board the {\em Chandra\/}
X-ray Observatory on 2002 May~22 UT with a 0.2-10.0~keV luminosity
$\sim10^{38}$~erg~s$^{-1}$ \citep{sr02}. A second observation three
weeks later showed that the X-ray luminosity had halved in that time
\citep{ems03}.

Radio monitoring of SN~2001ig with the Australia Telescope Compact
Array\footnote{The Australia Telescope is funded by the Commonwealth
of Australia for operation as a National Facility managed by CSIRO.}
(ATCA) commenced within a week of its discovery, and has continued on
a regular basis. In Section~\ref{s:data}, we present multi-frequency
radio flux data from the first 700~days. We describe our attempts to fit
the radio `light curves' with a circumstellar interaction model in
Section~\ref{s:model}, and discuss the deviations from this model in
more detail in Section~\ref{s:disc}. Our conclusions are presented in
Section~\ref{s:conc}.

\section[]{Radio monitoring}
\label{s:data}

Table~\ref{t:fluxes} contains the complete log of radio flux
measurements from the ATCA. Column~2 lists the days elapsed since
explosion, which is derived from the model fitting in
Section~\ref{s:model}. Total time on-source ranged from as little as
2~hours, up to a full 12~hour synthesis, but was typically 4--6
hours. Since the ATCA is capable of observing in 2 frequency bands
simultaneously, determining fluxes in 4 frequency bands on the same
day required time-sharing. Dual-frequency observations centered on
18.75 and 18.88~GHz and bandwidths of 128~MHz were carried out using a
prototype receiver system on just three ATCA antennas. The central
frequencies of the other bands are 8.640, 4.790, 2.496, and 1.376~GHz,
and the bandwidth is 128~MHz. From 2002~July~11 onwards, the S-band
central frequency was changed from 2.496 to 2.368~GHz, to reduce the
amount of in-band interference. The ATCA primary flux calibrator,
PKS~B1934-638 has been observed once per run, while observations of
the nearby source PKS~B2310-417 allow us to monitor and correct for
variations in gain and phase during each run.

\begin{table*}
\centering
\begin{minipage}{140mm}
\caption{SN~2001ig radio flux measurements with the ATCA.\label{t:fluxes}}
\begin{tabular}{lrrrrrc}
\hline
Date     & \multicolumn{1}{c}{Days since} & $S$(18.8~GHz) & $S$(8.6~GHz) & $S$(4.8~GHz) &
    $S$(2.4~GHz) & $S$(1.4~GHz) \\
(UT)     & 2001~Dec~3 UT & \multicolumn{1}{c}{(mJy)}     &
 \multicolumn{1}{c}{(mJy)} & \multicolumn{1}{c}{(mJy)}    &
 \multicolumn{1}{c}{(mJy)} & \multicolumn{1}{c}{(mJy)}    \\
\hline
15/12/01 &     12\sp  &      \md       & $2.1\pm0.3$  & $0.6\pm0.3$  &
        \nd       & \multicolumn{1}{l}{$<0.8$~~}     \\
18/12/01 &     15\sp  &      \md       & $3.9\pm1.6$  & $1.6\pm0.2$  &
        \nd       &     \nd       \\
22/12/01 &     19\sp  &      \md       & $8.7\pm0.8$  & $2.7\pm0.3$  &
        \nd       &     \nd       \\
26/12/01 &     23\sp  &      \md       & $15.0\pm3.0$ & $4.8\pm0.3$  &
        \nd       &     \nd       \\
31/12/01 &     28\sp  & $43\pm4$        & \nd          & $6.3\pm0.3$  &
 \multicolumn{1}{c}{$<4.6$} & \multicolumn{1}{l}{$<1.8$~~} \\
07/01/02 &     35\sp  &      \md       &    \nd        & $10.6\pm0.3$ &
        \nd       &     \nd       \\
10/01/02 &     38\sp  & $22\pm3$       & $23.6\pm4.1$ & $14.2\pm0.9$ &
         \nd       &     \nd       \\
15/01/02 &     43\sp  & $11\pm4$       & $18.9\pm5.6$ & $15.1\pm0.7$ &
        \nd       &     \nd       \\
20/01/02 &     48\sp  &      \md       &    \nd        &     \nd       &
 $5.3\pm0.3$     &     \nd       \\
02/02/02 &     61\sp  &      \md       & $9.9\pm2.2$  & $18.4\pm4.4$ &
        \nd       &     \nd       \\
17/02/02 &     76\sp  &      \md       & $7.0\pm1.1$  & $11.0\pm0.5$ &
        \nd       &     \nd       \\
26/02/02 &     85\sp  &      \md       &    \nd        &     \nd       &
 $11.5\pm0.3$    & \multicolumn{1}{c}{~$7.0\pm0.5$}  \\
17/03/02 &    104\sp  &      \md       & $10.7\pm2.0$ & $18.1\pm1.3$ &
        \nd       &     \nd       \\
19/03/02 &    106\sp  &      \md       &    \nd        &     \nd       &
 $25.1\pm1.3$    & $13.8\pm1.5$ \\
28/03/02 &    115\sp  &      \md       & $11.1\pm2.0$ & $21.9\pm1.0$ &
 $23.5\pm0.3$    & $14.2\pm0.6$ \\
09/04/02 &    127\sp  &      \md       & $12.6\pm0.7$ & $21.2\pm0.6$ &
        \nd       &     \nd       \\
22/04/02 &    140\sp  &      \md       & $12.7\pm0.9$ & $21.6\pm0.7$ &
        \nd       &     \nd       \\
12/05/02 &    160\sp  &      \md       & $12.4\pm0.5$ & $21.7\pm0.4$ &
 $30.3\pm1.6$    & $21.8\pm1.0$ \\
27/05/02 &    175\sp  &      \md       & $8.0\pm1.7$  & $15.6\pm1.3$ &
 $24.4\pm0.3$    & $21.5\pm0.7$ \\
09/06/02 &    188\sp  &      \md       & $6.2\pm1.7$  & $11.9\pm1.2$ &
 $19.7\pm1.0$    & $19.5\pm0.6$ \\
30/06/02 &    209\sp  &      \md       & $4.6\pm1.4$  & $8.9\pm1.0$  &
 $14.4\pm0.6$    & $19.5\pm0.8$ \\
11/07/02 &    220\sp  &      \md       & $4.0\pm1.2$  & $7.6\pm0.9$  &
 $14.9\pm0.4$    & $19.8\pm1.2$ \\
28/07/02 &    237\sp  &      \md       & $3.6\pm0.9$  & $7.1\pm0.7$  &
 $13.0\pm0.3$    & $18.2\pm0.5$ \\
11/08/02 &    251\sp  &      \md       & $3.3\pm1.0$  & $6.4\pm0.8$  &
 $12.7\pm0.3$    & $18.1\pm1.0$ \\
23/08/02 &    263\sp  &      \md       & $3.8\pm0.5$  & $6.7\pm0.3$  &
 $12.5\pm0.2$    & $17.8\pm0.9$ \\
31/08/02 &    271\sp  &      \md       & $3.7\pm0.3$  & $6.7\pm0.2$  &
 $12.5\pm0.2$    & $17.4\pm0.9$ \\
17/09/02 &    288\sp  &      \md       & $3.2\pm1.0$  & $6.6\pm0.7$  &
 $12.3\pm0.3$    & $~~16.4\pm0.6$\footnote{\citet{cr02} reported a flux
at 1.4 GHz of $11.7\pm1.5$~mJy for SN~2001ig on 2002 Sep 25.8 UT with the
Giant Meterwave Radio Telescope (GMRT). The source of this discrepancy has
since been traced to an elevation-dependent gain error in the GMRT
data \citep{cr03}.} \\
12/10/02 &    313\sp  &      \md       & $2.7\pm0.9$  & $6.7\pm0.4$  &
 $10.2\pm0.6$    & $15.4\pm0.8$ \\
22/11/02 &    354\sp  &      \md       & $2.5\pm0.3$  & $4.5\pm0.3$  &
 $8.0\pm0.3$     & $11.7\pm0.5$ \\
17/12/02 &    379\sp  &      \md       & $1.3\pm0.3$  & $3.6\pm0.3$  &
 $6.5\pm0.5$     & $11.5\pm1.1$ \\
05/02/03 &    429\sp  &      \md       & $1.8\pm0.4$  & $2.2\pm0.2$  &
 $6.1\pm0.5$     & $10.0\pm1.3$ \\
16/03/03 &    468\sp  &      \md       & $1.5\pm0.4$  & $2.6\pm0.3$  &
 $5.5\pm0.4$     & \multicolumn{1}{c}{~$8.8\pm0.4$} \\
20/05/03 &    533\sp  &      \md       & $1.2\pm0.5$  & $2.4\pm0.2$  &
 $4.5\pm0.2$     & \multicolumn{1}{c}{~$7.7\pm0.7$} \\
03/08/03 &    608\sp  &      \md       & $0.9\pm0.4$  & $2.0\pm0.2$  &
 $3.8\pm0.3$     & \multicolumn{1}{c}{~$6.3\pm0.4$} \\
07/11/03 &    704\sp  &      \md       & $0.8\pm0.3$  & $1.6\pm0.2$  &
 $3.1\pm0.3$     & \multicolumn{1}{c}{~$5.1\pm0.5$} \\
\hline
\end{tabular}
\end{minipage}
\end{table*}

\begin{figure*}
\includegraphics[width=15cm]{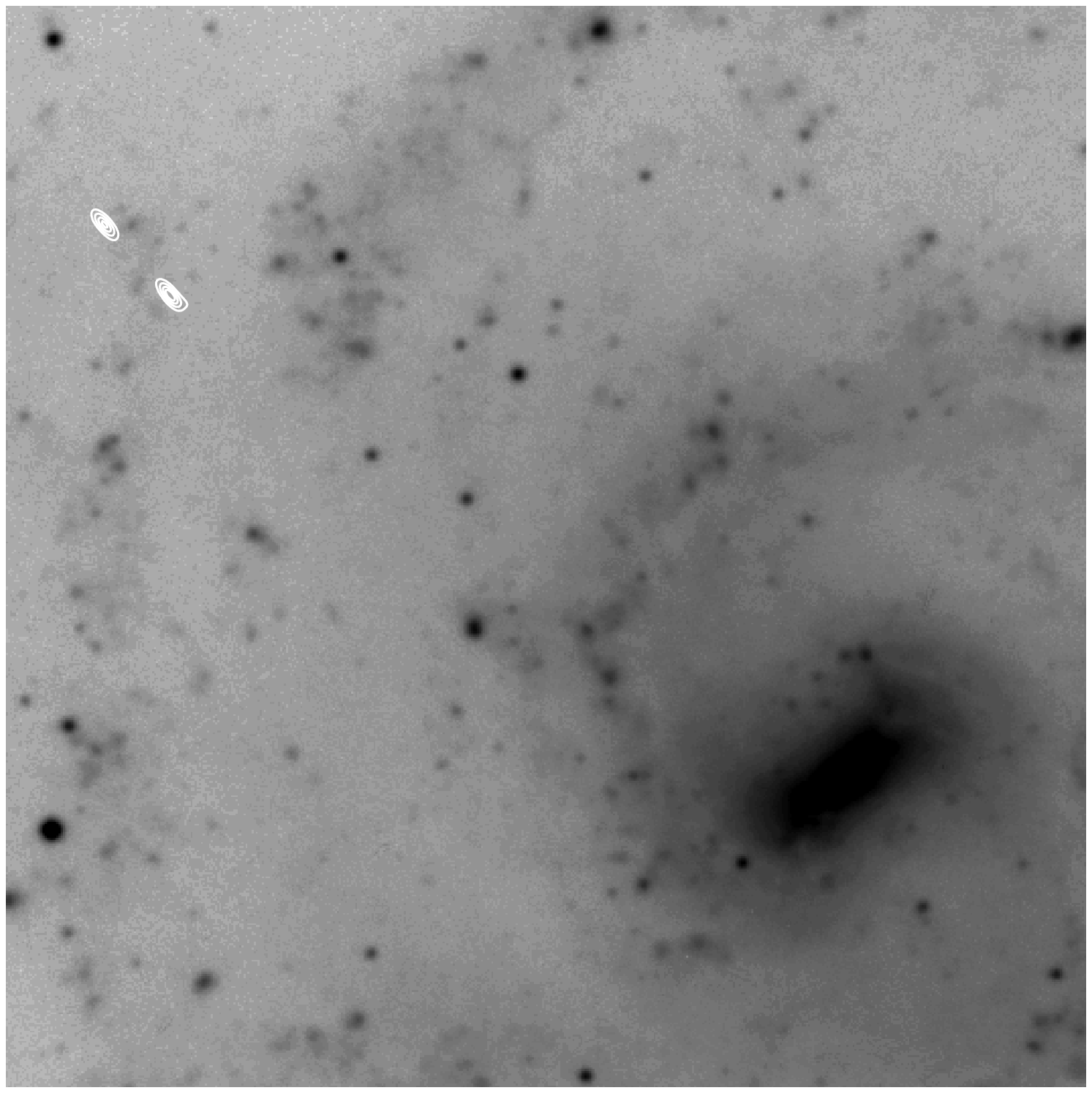}
  \caption{\label{f:c_on_v}Contours of 4.790~GHz radio emission, on a
  $V$-band image of NGC~7424. The radio observations were made with the
  ATCA on 2002 Feb~17~UT, and have a synthesised beamwidth of $3.8
  \times1.5$~arcsec. SN~2001ig is the upper-left of the two (unresolved)
  sources. The optical image is 3.5~arcmin on a side, obtained with the
  DFOSC instrument on the Danish 1.54-m telescope on La~Silla \citep{lr99},
  and made available through NED.}
\end{figure*}

The data for each observation and frequency have been edited and
calibrated using the {\sc miriad} software package. Rather than image
the visibility datasets, then clean them to some arbitrary level, the
UVFIT task is used instead in the visibility domain to fit
simultaneously a point source at the known location of SN~2001ig, as
well as a background source fortuitously located just 18.5~arcsec to
the southwest (Figure~\ref{f:c_on_v}). This source, located at
$\alpha=22^{\rm h} 57^{\rm m} 29.6^{\rm s}$, $\delta=-41^{\circ}
02^{\prime} 40^{\prime\prime}$ (J2000) does not appear in any radio
source catalogue, but was found to have the following fluxes as at
2001~Dec~15 UT: $S(8.640~{\rm GHz})=5.8$~mJy, $S(4.790~{\rm GHz})
=12.9$~mJy, $S(2.496~{\rm GHz})=26.4$~mJy, and $S(1.376~{\rm GHz})=
47.0$~mJy. After scaling the background source to these values at each
epoch, the SN~2001ig fluxes have been scaled accordingly.  The
uncertainties in Table~\ref{t:fluxes} include both the formal fitting
errors, and the possibility that this background source is
intrinsically variable (as is quite likely at the higher frequencies).
This technique of fitting the SN flux density in the visibility
domain, then bootstrapping to the flux of the adjacent source, was
crucial to recovering valid flux measurements from observations with
poor phase stability and/or limited hour-angle coverage.

The ATCA observations have been supplemented by a few early
observations with the Very Large Array (VLA)\footnote{The VLA
telescope of the National Radio Astronomy Observatory is operated by
Associated Universities, Inc. under a cooperative agreement with the
National Science Foundation}. The observing and data analysis
procedure follows that described in \citet{kw86}, and the results are
listed separately in Table~\ref{t:vla}. Owing to the low elevation of
SN~2001ig as observed from the VLA, and the compact configuration of
the VLA at the time, both the sensitivity and the resolution are
relatively poor. Nevertheless, these data prove to be important in
constraining the early evolution of SN~2001ig, particularly at the
very highest frequencies.

\begin{table*}
\centering
\begin{minipage}{140mm}
\caption{SN~2001ig radio flux measurements with the VLA.\label{t:vla}}
\begin{tabular}{lrrrrrr}
\hline
Date     & \multicolumn{1}{c}{Days since} & $S$(22.5~GHz) & $S$(15.0~GHz) &
    $S$(8.5~GHz) &   $S$(4.9~GHz) & $S$(1.4~GHz) \\
(UT)     & 2001~Dec~3 UT & \multicolumn{1}{c}{(mJy)}     &
 \multicolumn{1}{c}{(mJy)} & \multicolumn{1}{c}{(mJy)}    &
 \multicolumn{1}{c}{(mJy)} & \multicolumn{1}{c}{(mJy)}    \\
\hline
28/12/01 &     25\sp  & $37.2\pm5.6$   & $36.5\pm3.7$ & \nd          &
        \nd       & \nd           \\
07/01/02 &     35\sp  & $22.3\pm3.4$   & $30.9\pm3.1$ & \nd          &
        \nd       &     \nd       \\
10/01/02 &     38\sp  & $16.6\pm2.5$   & $25.1\pm2.6$ & \nd          &
        \nd       &     \nd       \\
13/01/02 &     41\sp  & $15.8\pm2.4$   & $20.5\pm2.1$ & $23.5\pm1.8$ &
        \nd       &     \nd       \\
17/01/02 &     45\sp  &      \nd       &    \nd       &     \nd      &
        \nd       & $1.6\pm0.4$   \\
27/01/02 &     55\sp  &      \nd       &    \nd       &     \nd      &
 $9.8\pm0.7$      & $3.2\pm0.3$   \\
21/03/02 &    108\sp  &      \nd       &    \nd       & $10.5\pm0.5$ &
 $18.6\pm0.9$    & $12.6\pm0.8$   \\
\hline
\end{tabular}
\end{minipage}
\end{table*}

\section[]{Radio light curves}
\label{s:rlc}

\begin{figure*}
\rotatebox{270}{\includegraphics[width=12cm]{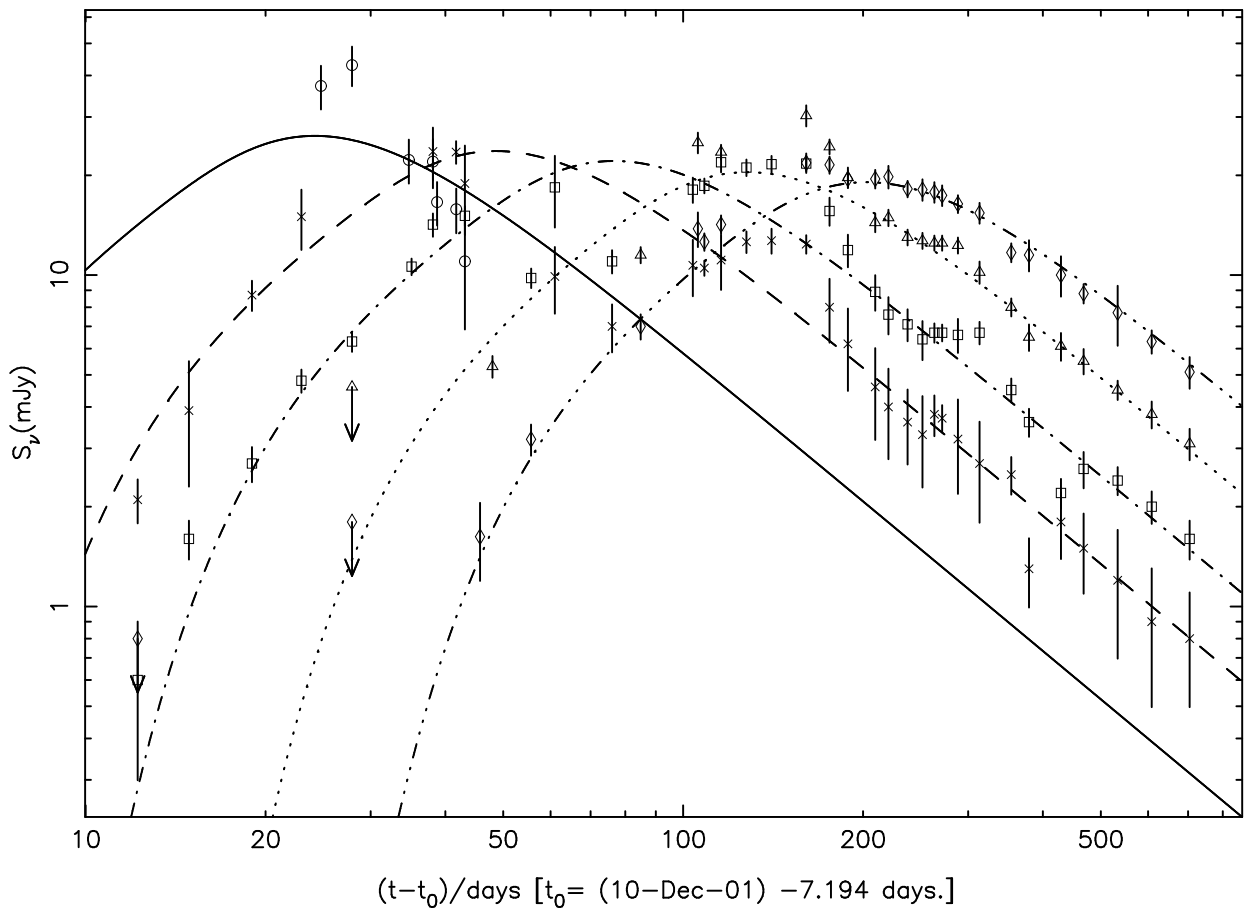}}
  \caption{\label{f:kurt}SN 2001ig at radio frequencies of 22.5/18.8~GHz
  ({\em circles, thick solid line}), 8.6/8.5~GHz ({\em crosses, dashed line}),
  4.9/4.8~GHz ({\em squares, dash-dotted line}), 2.4~GHz ({\em triangles,
  dotted line}), and 1.4~GHz ({\em diamonds, dash-triple dotted line}).
  The curves are a model fit to the data, as described in the text.}
\end{figure*}

\begin{figure*}
\rotatebox{270}{\includegraphics[width=10cm]{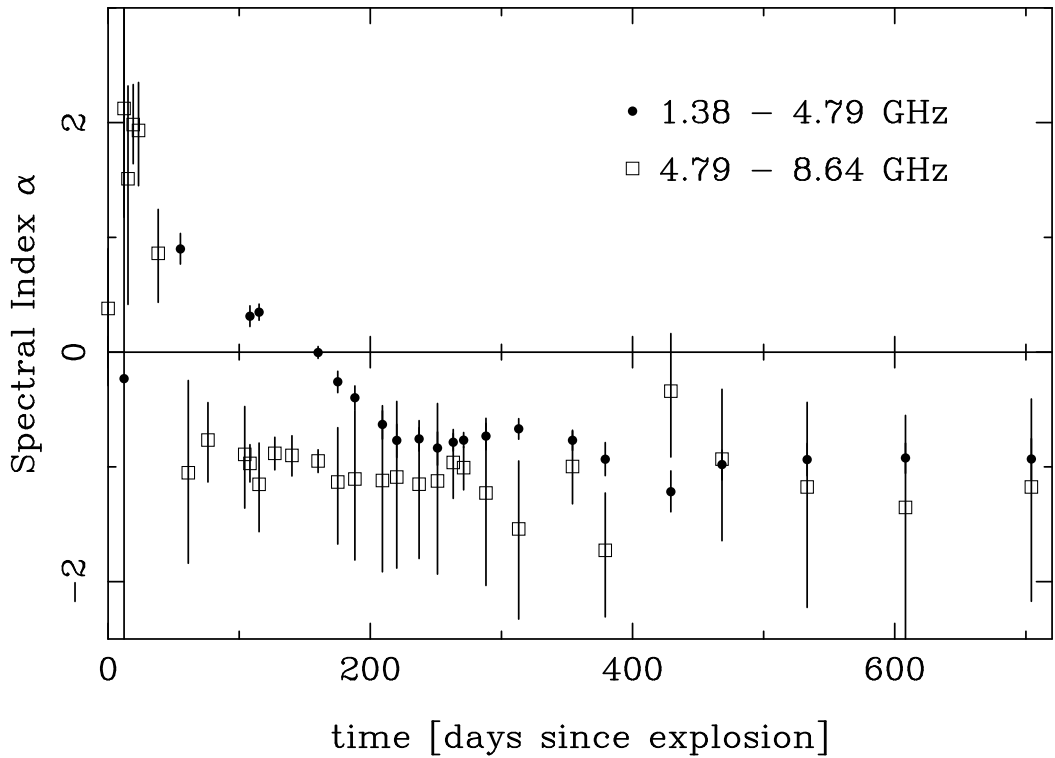}}
  \caption{\label{f:spix}Evolution of spectral index $\alpha$ for SN~2001ig,
  plotted linearly as a function of time, between 1.38 and 4.8/4.9~GHz
  ({\em solid circles}); and between 4.8/4.9 and 8.64~GHz
  ({\em open squares}).}
\end{figure*}

The ATCA and VLA radio data is plotted in Figure~\ref{f:kurt}. The
data at 15.0~GHz are not shown here, to reduce confusion with the other
high-frequency points, but is incorporated in the model fitting of
Section~\ref{s:model}. The time evolution of the spectral index
$\alpha$ (where flux $S \propto \nu^{+\alpha}$) between simultaneous
positive detections at 1.4 and 4.8/4.9~GHz, and between 4.8/4.9 and
8.6~GHz, is plotted in Figure~\ref{f:spix}.

The radio ``light curve'' of a Type~II supernova can be broadly
divided into three phases: first, there is a rapid turn-on with a
steep spectral index ($\alpha>2$, so the SN is brightest at the higher
frequencies), due to a decrease in the line-of-sight absorption. After
some weeks or months have elapsed, the flux reaches a peak, turning
over first at the highest frequencies. Eventually, the SN begins to
fade steadily, and at the same rate at all frequencies, in the
optically-thin phase.

Although broadly consistent with this picture, the radio light curve
of SN~2001ig displays significant departures from a smooth turnover
and decline, which are most pronounced at 8.64 and 4.79~GHz. At around
day~80, the flux at these frequencies reversed its initial decline,
and by day~130 had almost doubled. The flux remained almost constant
for a period of $\sim50$~days, before resuming its decline at close to
the original rate.  Near day~250, the decline was again temporarily
interrupted for another 50~days. There are indications of perhaps one
more bump after day~450, but by this stage the SN has faded to the
milli-Jansky level where any variations are comparable to the
measurement uncertainties. The deviations are less pronounced, but
still evident at 2.50 and 1.38~GHz.

\section[]{Model fits}
\label{s:model}

The general properties of supernova radio light curves as outlined
above are quite well represented by a modified version of the
``minishell'' model of \citet{che82}, and have been successfully
parameterised for more than a dozen RSNe (see Table~2 of
\citet{kw02}). Radio synchrotron emission is produced when the SN
shock wave ploughs into an unusually dense circumstellar medium
(CSM). Following the notation of \citet{kw02} and \citet{sw03}, we
model the multi-frequency evolution as:

\begin{eqnarray} 
S {\rm (mJy)} = K_1 {\left({\nu} \over {\rm 5~GHz}\right)^{\alpha}}
{\left({t - t_0} \over {\rm 1~day}\right)^{\beta}} e^{-{\tau_{\rm external}}} 
 \nonumber \\
\times
{\left({1 -e^{-{\tau_{\rm CSM_{clumps}}}} \over {\tau_{\rm CSM_{clumps}}}}
\right)} {\left({1 - e^{-{\tau_{\rm internal}}}} \over {\tau_{\rm internal}}
\right)}
\end{eqnarray}

\noindent
with 

\begin{equation}
\tau_{\rm external} = \tau_{\rm CSM_{homog}} + \tau_{\rm distant}
\end{equation}

\noindent
where

\begin{equation} 
\tau_{\rm CSM_{homog}} = K_2 {\left({\nu} \over {\rm 5~GHz}\right)^
{-2.1}}
{\left({t - t_0} \over {\rm 1~day}\right)^{\delta}},
\end{equation}

\begin{equation} 
\tau_{\rm distant} = K_4 {\left({\nu} \over {\rm 5~GHz}\right)^{-2.1}},
\end{equation}

\noindent
and

\begin{equation} 
\tau_{\rm CSM_{clumps}} = K_3 {\left({\nu} \over {\rm 5~GHz}\right)^{-2.1}}
{\left({t - t_0} \over {\rm 1~day}\right)^{\delta^{\prime}}},
\end{equation}

\noindent with the various $K$ terms representing the flux density
($K_1$), the attenuation by a homogeneous absorbing medium ($K_2$,
$K_4$), and by a clumpy/filamentary medium ($K_3$), at a frequency of
5~GHz one day after the explosion date $t_0$. The $\tau_{\rm
CSM_{homog}}$ and $\tau_{\rm CSM_{clumps}}$ absorption arises in the
circumstellar medium external to the blast wave, while $\tau_{\rm
distant}$ is a time-independent absorption produced by e.g., a
foreground H\,{\sc ii}~region or more distant parts of the CSM
unaffected by the shock wave. The spectral index is $\alpha$; $\beta$
gives the rate of decline in the optically-thin phase; and $\delta$
and $\delta^{\prime}$ describe the time dependence of the optical
depths in the local homogeneous, and clumpy/filamentary CSM,
respectively (see \citet{kw02} and \citet{sw03} for a detailed account
of how these parameters are related). For lack of sufficient
high-frequency data prior to the turnover to constrain it, we adopt
${\tau_{\rm internal}}=0$.

In order to assess the gross properties of SN~2001ig in comparison
with other Type~IIb RSNe, we have fit this standard model to the data
in Tables~\ref{t:fluxes} and \ref{t:vla}, but excluding days~48--70,
and days~110--190. Thus, the model fit is constrained primarily by the
rise at early times, the region of the high-frequency turnover, and by
the late-time decay. The actual date of explosion $t_0$ is found to be
2001~Dec~3~UT, one week prior to discovery. The full set of model
parameters which yields the minimum reduced $\chi^{2}$ is given in
Table~\ref{t:fits}, and the model curves are plotted in
Fig.~\ref{f:kurt}. For comparison, we show in Table~\ref{t:fits} the
equivalent parameters for two other well-sampled Type~IIb RSNe:
SN~1993J \citep{vd04} in M81, and SN~2001gd \citep{sto03} in NGC
5033. Note that we have fixed the value of $\delta$ to be $(\alpha -
\beta -3)$, as in the \citet{che82} model for expansion into a CSM
with density decreasing as $r^{-2}$.

\begin{table*}
\centering
\begin{minipage}{140mm}
\caption{Comparison of radio light curve model parameters.\label{t:fits}}
\begin{tabular}{cccc}
\hline
Parameter    &     SN 2001ig      &      SN 1993J      &    SN 2001gd  \\
\hline
$K_1$        & $2.71\times10^{4}$ & $1.36\times10^{4}$ & $1.49\times10^{3}$ \\
$\alpha$     &     $-1.06$        &     $-1.05$        &     $-1.38$        \\
$\beta$      &     $-1.50$        &     $-0.88$        &     $-0.96$        \\
$K_2$        & $1.38\times10^{3}$ & $9.14\times10^{2}$ & $3.25\times10^{6}$ \\
$\delta$     &     $-2.56$        &     $-1.88$        &      --            \\
$K_3$        & $1.47\times10^{5}$ & $8.33\times10^{4}$ & $1.05\times10^{3}$ \\
$\delta^{\prime}$ & $-2.69$       &     $-2.26$        &     $-1.27$        \\
$K_4$        &        0.0        & $2.76\times10^{-3}$ &      --            \\
Time to $L_{\rm 5\ GHz \ peak}$ (days) & 74 & 167      &      173           \\
$L_{\rm 5\ GHz \ peak}$ (erg s$^{-1}$ Hz$^{-1}$) &
            $3.5\times10^{27}$ & $1.4\times10^{27}$   & $2.9\times10^{27}$ \\
Mass-loss rate (${\rm M_\odot}$ yr$^{-1}$) &
      $(2.2\pm0.5)\times10^{-5}$  & $2.1\times10^{-5}$ & $3.0\times10^{-5}$ \\
\hline
\end{tabular}
\end{minipage}
\end{table*}

The spectral index $\alpha$ of SN~2001ig is virtually identical to
SN~1993J, but less steep than SN~2001gd. However, the rate of decline
$\beta$ is much steeper in SN~2001ig than in either of the other Type
IIb's, and the time to reach peak 5~GHz flux is also much shorter. In
this respect, SN~2001ig has behaved more like a Type Ib/c SN than most
``normal'' Type~II SNe. The interpolated peak 5~GHz luminosity would
be about twice that attained by SN~1993J, though in practice SN~2001ig
was near a local minimum in the flux at that time, and the actual peak
was not reached for another 40~days.

Using the methodology outlined in \citet{kw02} and \citet{sw03}, we
can derive an estimate of the progenitor's mass-loss rate, based on
its radio absorption properties. Substituting our model fit results
above into their equation~11, and assuming both $\tau_{\rm
CSM_{homog}}$ and $\tau_{\rm CSM_{clumps}}$ contribute to the
absorption\footnote{Case~2 in the notation of \citet{kw02}. Note that
there is a misprint in that section, namely that in the limit of
$\tau_{\rm CSM_{clumps}} \rightarrow 0$, then $\langle \tau^{0.5}_{\rm
eff} \rangle \rightarrow \tau^{0.5}_{\rm CSM_{homog}}$ in
equation~13, and not $\tau_{\rm CSM_{homog}}$. We prefer the
term ``homogeneous'' here over ``uniform'', since the latter could
give the misleading impression of no density gradient at all, whereas
an $r^{-2}$ dependence of density is implicit.}, we find that

\begin{displaymath}
\frac{\rm \dot M/(M_{\odot}\ {\rm yr}^{-1})}{w /{\rm (10\ km\ s^{-1})}}
= (2.2 \pm 0.5) \times 10^{-5}
\end{displaymath}

\noindent
where $w$ is the mass-loss wind velocity, and the ejecta velocity as
measured from the earliest optical spectra \citep{cp01,clo02} is in
the range 15,000 -- 20,000~km~s$^{-1}$. Clearly, this is only an
average value, subject to major variations discussed in the next
section, but is in the same domain as the mass-loss rates derived
similarly for SN~1993J and SN~2001gd (Table~\ref{t:fits}). Though
generally less well-constrained, mass-loss rates may also be estimated
directly from the radio emission properties, relying only on the peak
5~GHz luminosity, and the time taken to reach that peak. Equations 17
and 18 of \citet{kw02} give ${\rm \dot M}/{w} = 1.5 \times 10^{-5}$
and $3.5 \times 10^{-5}$ for the average Type~Ib/c SN and Type II SN,
respectively. Thus, the mass-loss rate calculations are in good
agreement, with SN~2001ig intermediate between the expected rates for
Type Ib/c and Type II SNe, consistent with its Type IIb classification.

\section[]{Discussion}
\label{s:disc}

\begin{figure*}
\rotatebox{270}{\includegraphics[width=12cm]{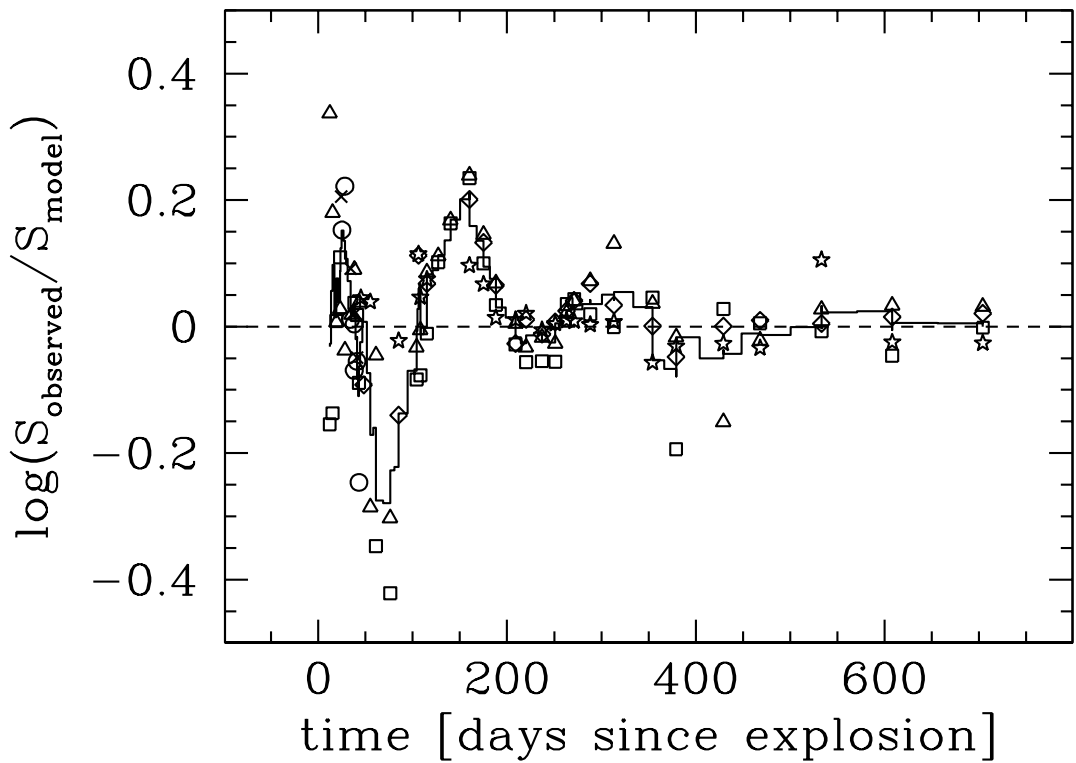}}
  \caption{\label{f:devs}Deviations of the observed flux (on a log scale)
  about the best-fit model, plotted linearly as a function of time. The
  symbols represent frequencies of 22.5/18.8~GHz ({\em circles}), 15.0~GHz
  ({\em crosses}), 8.6/8.5~GHz ({\em squares}), 4.9/4.8~GHz ({\em triangles}),
  2.4~GHz ({\em diamonds}), and 1.4~GHz ({\em stars}), while the solid line
  is a smoothed interpolation, as described in the text.}
\end{figure*}

In Figure~\ref{f:devs} we have plotted the deviations of the observed
flux density, from the best-fit model curves as shown in
Figure~\ref{f:kurt}. The solid line in this figure is a 4-point boxcar
average of the mean deviation over all frequencies at each epoch,
which serves to emphasise the quasi-damped harmonic nature of the
deviations, having a period near 150~days, and peak intensity
declining with time (i.e., with increasing distance from the star).
The r.m.s. deviation of the actual data from the smoothed
interpolation is less than 1/3 of the amplitude of the observed
modulation. Since the fractional amplitude of these deviations is
virtually identical at each frequency, the evolution of the spectral
index (Fig~\ref{f:spix}) in the optically-thin phase appears
relatively unaffected. We take this as evidence that the bumps and
dips in the radio light curve primarily reflect abrupt modulations in
the CSM density structure, rather than optical depth effects (though
optical depth is tied to CSM density to some degree). We now consider
ways in which such a structured CSM may have been laid down late in
the life of the progenitor of SN~2001ig.

Before doing so, we need to examine the effects of a change in CSM
density on the velocity of the expansion, as well as on the radio
emission. The blastwave radius increases with time as $r \propto t^{m}$,
where $m=1$ for no deceleration. Since $m = - \delta / 3$ in the
\citet{che82} model, then $m= 0.85$ for SN~2001ig, implying
significant deceleration in the surrounding CSM. The radio luminosity
is related to the average CSM density ($\rho_{\rm CSM} \propto
{\rm \dot M} / w$) via

\begin{equation}
L \propto \left( \frac{\rm \dot M}{w} \right)^{(\gamma - 7 + 12m)/4}
\end{equation}

\noindent
\citep{che82} so that for SN~2001ig, $L \propto ({\rm \dot
M}/w)^{1.6}$.  Consequently, a doubling in the CSM density will cause
the radio emission to rise by a factor of three.

\subsection{Episodic mass-loss from a single progenitor\label{s:tp}}

The observed transition in SN~2001ig from a Type~II optical spectrum
with H lines, to a Type Ib/c spectrum without, argues for the ejection
of a significant fraction of the red-giant envelope. Fig.~\ref{f:devs}
indicates strong excesses of observed flux at $t\sim150$~days, and
$t\sim300$~days (with a net flux excess still at 500--600~days),
hinting at a possible periodicity of 150~days in CSM density
enhancements. If these density enhancements (by factors of 30\% and
15\% respectively) correspond to discrete shells of material expelled
by the red supergiant, then the spacing between these shells is given
by $R_{\rm sh} = v_{\rm exp} t^{m}$. Given an initial ejecta expansion
velocity $v_{\rm exp}$ of 15,000 -- 20,000~km~s$^{-1}$
\citep{cp01,clo02}, an elapsed time $t$ of 150~days, and deceleration
given above, then $R_{\rm sh} = (6.3 \pm 0.9) \times
10^{-4}$~pc. Assuming the shells have been expanding at the wind
velocity $w = 10-20$~km~s$^{-1}$, then the period between successive
mass-loss episodes is $T\sim20-60$~years.

This is significantly longer than the timescales normally associated
with stellar pulsations ($T\sim$ few hundred days) in asymptotic giant
branch (AGB) stars, but is comparable with the expected
$10^{2}-10^{3}$~yr intervals between thermal pulses (C/He shell
flashes) in $5-10$~M$_{\odot}$ AGB stars \citep{ir83}. Computations by
\citet{bp75} predict an interflash period of 40~yrs when the
(hydrogen-exhausted) core mass is nearly 1.3~M$_{\odot}$.
This is close to the maximum core mass possible (depending on
composition; \citet{bi80}) in stars which will finish up as a white
dwarf, rather than undergo a supernova explosion. Put another way,
SN~2001ig may span the gap between the most extreme mass-losing AGB
stars, and the least-massive supernova progenitors. Such a massive AGB
star would also be expected to undergo multiple convective
``dredge-up'' phases leading to modified surface abundances, with N
enhanced primarily at the expense of C \citep{bi79}. Such changes may
be apparent in the optical spectra \citep{sm04}.

There are already several good examples within our own Galaxy for this
type of quasi-periodic mass-loss from AGB and post-AGB/proto-planetary
nebula objects. The nearby carbon star IRC$+10216$ shows multiple
nested shells spaced $5-20$~arcsec apart, corresponding to ejection
timescales of $200-800$~years \citep{mh00}. The ``Egg Nebula''
(CRL2688; \citealt{rs98}) and IRAS~1710--3224 \citep*{sk98} are two
bipolar proto-planetary nebulae which display concentric arcs with
spacings on similar timescales.

\subsection{Binary model}
\label{s:binary}

Among the other RSNe studied to date, only SN~1979C shows anywhere
near as much systematic variation in its optically-thin decline phase
as SN~2001ig. Indeed, for the first decade or so, the late-time radio
light curve of SN~1979C could be quite well represented by a
sinusoidal modulation of the flux, with a period of 1575~days
\citep{kw92}. More recently however, these regular variations in
SN~1979C have ceased, and the light curve has flattened out
\citep{mm00}. The implied modulation period of the mass-loss rate is
$T\sim4000$~years, a factor of 100 longer than that computed above for
SN~2001ig. On the basis of evidence that the progenitor of SN~1979C
had an initial mass of at least 16~M$_{\odot}$, \citet{kw92} argued
against the thermal pulse scenario described in Section~\ref{s:tp}, as
the interflash period for such a massive star and its resulting core
would be $\ll 4000$~years. Instead, they proposed modulation of the
progenitor wind due to eccentric orbital motion about a massive binary
companion as the cause of the periodicity in the radio light curve of
SN~1979C.

The particular binary scenario they presented had the red supergiant
progenitor and a 10~M$_{\odot}$ B1 dwarf orbiting their common
barycentre, with a 4000~year period. For a purely circular orbit, the
progenitor's orbital motion is a sizable fraction of the wind
velocity, resulting in a spiral (or pinwheel-like) density structure
being imprinted on the (otherwise uniform) mass-loss CSM in the
orbital plane. This by itself would not lead to any periodic variation
in the CSM density swept up by the SN shock wave. If instead the orbit
was eccentric ($e=0.5$ say), the acceleration of the progenitor near
periastron every 4000~years would cause an additional pile-up of wind
material which may then account for the observed periodicity in the
radio emission. \citet{sp96} performed full hydrodynamical simulations
of this, taking into account shocks generated in the wind, the
gravitational influence of the companion on the wind, and light-travel
time effects. Their simulations produced pronounced, but asymmetric
spiral shock patterns, particularly when a polytropic equation of
state is assumed. They also demonstrated that the amplitude of the
modulations in the radio light curve tend to decrease as our view of
the orbital plane goes from edge-on, to face-on. This may partly
explain why such periodic variations as seen in SN~1979C and in
SN~2001ig are comparatively rare, since not only must the mass-loss
rate be modulated by the right kind of binary orbit parameters, but we
must also then be fortunate enough to view it from close to edge-on.

Similar hydrodynamical simulations, but in three dimensions, of
detached binary systems comprising a 1.5~M$_{\odot}$ AGB star and a
secondary of $0.25-2.0$~M$_{\odot}$ were presented by \citet{mm99},
specifically targeted at reproducing the observed structures in
IRC$+10216$, CRL2688, etc., mentioned in Section~\ref{s:tp}.
Interestingly, for sufficiently large binary separations ($>$10~AU)
and wind velocities ($w>15$~km~s$^{-1}$), the spiral shock structure
extends to high latitudes, and the resulting ``spiral onion shell''
structure seen in cross-section could resemble the shells seen in
PPNe, and implied in SN~2001ig. In an alternative binary scenario,
\citet*{hrs97} postulate that it is the close passage of the secondary
star effectively ``choking off'' uniform mass-loss from an AGB star
having an extended atmosphere, rather than any enhancements in the
mass-loss rate itself, which gives rise to these shells.

Perhaps the best direct evidence for the existence of binary-generated
spiral shocks comes from high-resolution observations of dusty
Wolf-Rayet (W-R) stars. \citet*{tmd99} and \citet*{mtd99} used the
technique of aperture-masking interferometry on the Keck~I telescope
to image structure at better than 50~mas resolution in the $K$-band
around WR~104 and WR~98a. In both cases, they found pinwheel-shaped
nebulae wrapping almost entirely around the central source out to
distances of 150--300~AU. Their model has an OB-type companion
orbiting the W-R star, with dust formation taking place in the wake of
the interface region between their colliding stellar winds. The
combination of orbital motion and wind-driven radial motion results in
a ``lawn sprinkler'' effect, and the resultant dusty
spirals. Hydrodynamic models in 3D of these colliding wind binary
systems by \citet{wf03} show the effects of varying the orbital
eccentricity.

Remarkably, the characteristic radial scale for density enhancements
implied by SN~2001ig's radio light curve ($R_{\rm sh} = 0.0006$~pc;
Sect.~\ref{s:tp}) is an almost perfect match to the typical scale of
one full rotation of these pinwheel nebulae: 50--100~milliarcsec at
$D\sim2~{\rm kpc} \rightarrow R = 0.0005-0.001$~pc. Thus, SN~2001ig
may represent the obliteration of just such a pinwheel nebula.
Further support for this scenario comes from the flux excess (at least
60\%, though this is a lower limit due to optical depth effects) in
the first 50~days over that expected from a simple $r^{-2}$ CSM
density profile, as shown in Fig.~\ref{f:devs}.  Coupled with the
deceleration mentioned previously, this would tend to favour a
centrally-condensed additional CSM component such as a pinwheel
nebula, rather than concentric mass-loss shells.

\subsection{SN~2001ig and the link between Type II and Type Ib/c SNe}

It is interesting to note that the apparent requirements to produce
such a pinwheel nebula, i.e., a close binary system composed of two
massive stars in which the primary is a W-R star,
are similar to those invoked by stellar evolution theory to explain
the origin of Type Ib/c SNe. The peculiar spectral evolution and
optical light curve behaviour of SN~1993J and SN~1987K
(Sect.~\ref{s:intro}) has been attributed to the explosion of a
hydrogen-poor, helium-rich progenitor \citep{swa93}. A large fraction
of the progenitor's original hydrogen envelope must have been shed
prior to core collapse, either through a strong stellar wind from a
single massive (25--30~M$_{\odot}$) star \citep*{hld93}; or more
likely, via mass transfer from an intermediate-mass
(10--15~M$_{\odot}$) star in a binary system
\citep{nom93,pod93,utr94,vhf96}. Models for the evolution of massive stars
in close binaries (e.g., \citealt*{pjh92}; \citealt*{wlw95}) produce
helium stars. The suggestion is that stars which have lost some or all
of their hydrogen would be the WN class of W-R stars, and explode as
Type~Ib SNe; while those which lose their helium layer as well would
be the WC or WO classes of W-R stars, and explode as Type~Ic SNe
(\citealt{har87}, \citealt*{fmh93}).

At least 40\% of solar neighbourhood W-R stars are in binary systems
with hot companions \citep{mof86,vdh88}, and the fraction may be even
higher in low-metallicity environments \citep{ds95}, such as the
outskirts of NGC~7424. The extent of mass transfer, and thus the
end-products of the binary system, depends on the evolutionary stage
of the primary at the time mass transfer commences. As shown by
\citet{pjh92} and \citet{pn97}, Case~C mass transfer (which takes
place after the core helium-burning phase) in systems with large
eccentricity and orbital periods of a few years can be episodic,
occurring mainly near periastron, just as outlined in
Sect.~\ref{s:binary}. We propose that SN~2001ig may well have
undergone just such a phase, without actually sharing a common
envelope with a companion. The implied orbital period is given by one
complete winding of the pinwheel nebula, which is simply the ratio of
$R_{\rm sh} \sim 0.0006$~pc divided by the terminal wind velocity of a
WN star ($\sim2000$~km~s$^{-1}$; \citealt{ac87}), or $T\sim100$~days,
consistent with this scenario.

In this context, the highly-modulated radio light curves for SN~2001ig
may represent some of the best evidence yet for a link between SNe of
Type~IIb and Type~Ib/c, in that SN~2001ig evolved optically like a
Type~IIb, but has the radio characteristics that should be expected
for a Type~Ib/c SN originating in a W-R + OB binary system viewed
nearly edge-on. A testable prediction from this scenario is that the
companion star (which by virtue of mass accretion may be even brighter
than the W-R progenitor of SN~2001ig) should eventually become visible
against the fading optical remnant, as has been postulated but not yet
observed for SN~1993J \citep{pod93,fmh93}.

Besides its apparent association with the gamma-ray burst GRB~980425,
SN~1998bw is also notable for showing bumps and dips in its radio
light curves not dissimilar to those seen in SN~2001ig
\citep*{wpm01}. These deviations, while exhibiting no clear
periodicity in the first 1000~days, were less conspicuous at low
frequencies, just as in SN~2001ig. This kind of behaviour is likely
due to the CSM still being optically thick to low frequencies at
relatively late times, and therefore only emission originating in the
near-side of the CSM can be seen; whereas at high frequencies,
emission from the entire CSM is visible. SN~1998bw had the spectral
characteristics primarily of a Type~Ic event, with a probable W-R
progenitor \citep{iwa98}. Unfortunately, barely a half-dozen Type Ib/c
SNe have been detected or studied in the radio, so it is too early to
tell whether such modulated radio light curves may be the clue to
a common massive-binary origin for Type IIb and Type~Ib/c~SNe, and
possibly some GRBs.

\section{Conclusions}
\label{s:conc}

By compiling one of the most complete multi-frequency radio datasets
ever collected for any supernova, we have been fortunate to witness
regular modulations in the radio light curves of the Type IIb
Supernova~2001ig. The time taken to reach peak luminosity at 5~GHz,
the rate of decline since then, and the derived mass-loss rates
prior to explosion are all intermediate between those of Type Ib/c
and ``normal'' Type~II SNe. We find the light curve modulations to
recur on a timescale of $\sim150$ days, and have shown them to be
true density modulations in the CSM, and not optical depth effects.
Allowing for the deceleration of the ejecta, these density enhancements
are spaced 0.6~milli-parsecs (or 130~AU) apart.

While we cannot totally exclude the possibility that these density
enhancements represent mass-loss shells from the thermal-pulsing phase
of a single AGB star progenitor, we find the weight of evidence
supports a stellar wind, modulated by motion in an eccentric binary
system, as their source. As had been suggested previously for
SN~1979C, the combination of a massive binary companion causing a
pile-up of mass-loss during periastron, and a favourable viewing
angle, can result in just the kind of periodic density variations
observed in SN~2001ig. Recent near-IR interferometric observations of
the anticipated ``pinwheel'' dust nebulae in systems comprising a
Wolf-Rayet star and a hot massive companion lend weight to this
scenario, especially as the observed size scales match those required
for SN~2001ig. Finally, as optical spectroscopy has recently been
leading to the conclusion that Type Ib/c SNe are the product of a
core-collapse event in the W-R component of such systems (and Type IIb
SNe being those caught early enough to reveal the last traces of their
lost hydrogen envelope), we believe these radio observations of SN~2001ig
provide the ``missing link'' between the pinwheel nebulae observed in
Galactic W-R binary systems, and their eventual fate in Type IIb or
Type Ib/c supernovae.

\section*{Acknowledgments}

We are grateful to the staff of the Paul Wild Observatory, including
numerous volunteer Duty Astronomers, for assisting us in carrying out
the majority of these observations remotely from ATNF Epping.  This
research has made use of NASA's Astrophysics Data System Bibliographic
Services (ADS), as well as the NASA/IPAC Extragalactic Database (NED)
which is operated by the Jet Propulsion Laboratory, California
Institute of Technology, under contract with the National Aeronautics
and Space Administration. We acknowledge fruitful discussions with Tim
Gledhill and Roger Chevalier. KWW wishes to thank the Office of Naval
Research (ONR) for the 6.1 funding supporting his research.


\bsp

\label{lastpage}

\end{document}